\documentclass[%
 reprint,
 aps,
 prl,
 superscriptaddress,
 amsmath,amssymb,
 floatfix
]{revtex4-2}

\usepackage{transparent}
\usepackage{amsmath}
\usepackage{graphicx}
\usepackage{amssymb}
\usepackage{epstopdf}
\usepackage{braket}
\usepackage{mathtools}
\usepackage{bm} 
\usepackage{longtable}
\usepackage{multirow}
\usepackage{enumerate}
\usepackage{hyperref}
\usepackage{xcolor}
\usepackage{physics}
\usepackage{array}
\usepackage{lipsum}
\usepackage{tabularx}
\usepackage{printlen}
\usepackage{textcomp}
\usepackage{sidecap}
\hypersetup{colorlinks=true,       
            linkcolor=blue,        
            citecolor=blue,        
            filecolor=blue,        
            urlcolor=blue,         
            runcolor=blue
}
\usepackage{svg}
\usepackage{tikz,lmodern}
\usepackage{blindtext}

\usepackage{array}

\begin{document}

\title{Coherent Control of Single-Photon Emission in Broadband Plasmonic Resonators}

\author{Athul S. Rema}

\author{Adri\'an E. Rubio López}

\affiliation{Department of Physics, Universidad de Santiago de Chile, Av. Victor Jara 3493, Santiago,  9170124, Chile.}

\author{Felipe Herrera}
\affiliation{Department of Physics, Universidad de Santiago de Chile, Av. Victor Jara 3493, Santiago,  9170124, Chile.}
\affiliation{Millennium Institute for Research in Optics, Concepci\'on, Chile.}

\date{\today}

\begin{abstract}
Encoding phase information with quantum emitters in metallic optical nanocavities is challenging due to ultrafast relaxation induced by the broadband nature of the plasmonic near field. Using macroscopic quantum electrodynamics theory, we demonstrate that an individual emitter strongly coupled to a plasmonic nanocavity field that is driven by a pulsed laser with frequency comb spectrum undergoes spontaneous emission of well-resolved one-photon pulses whose amplitude scales quadratically with the number of frequency comb components due to coherent accumulation, and ultrashort non-exponential temporal profiles with duration set by the light-matter coupling strength. Conditions are discussed for exponentially suppressing the probability of two or more emission events within the lifetime of the coupled emitter-nanocavity system. The spectral signatures of the optimal one-photon wavepackets are discussed. This work paves the way for the development of ultrafast laser-triggered single-photon emitters using atoms and molecules in optical nanocavities.
\end{abstract}

\maketitle

Controlling single-photon emission is a long-standing goal in photonic quantum technology~\cite{Kimble2008,OBrien2009,Wehner2018,Tudela2024}. Solid-state emitters such as quantum dots, molecules, and color centers can deliver high-purity photons in engineered photonic environments~\cite{Senellart2017,Aharonovich2016,Toninelli2021,Sipahigil2016}. Control is usually achieved by driving emitters directly with spectrally shaped electromagnetic fields, building on schemes designed for atomic and molecular emitters in free space under weak light-matter coupling conditions~\cite{Lodahl2015,Weiner2000,Brif2010,Brixner2001,Rabitz2000}. Such schemes are difficult to extrapolate to optical nanocavities due to their highly absorptive and dispersive nature~\cite{Koenderink2010,Khurgin2015,Hoang2016,Bitton2022,Shahbazyan2022}, which can result in significant modifications of the amplitude and phase profiles a far-field laser field when reaching the nanoscale ~\cite{Aeschlimann2007,Volpe2009,vanHulst2014,Kojima2016,Lu2024}. Metallic nanocavities based on localized plasmonic resonances are particularly promising light-matter interfaces because their extreme mode confinement enables resonant strong coupling with individual quantum emitters at room temperature~\cite{Torma2015,Chikkaraddy2016,Franke2019}. However, stronger emitter-nanocavity interaction also implies strong reduction of the emitter lifetime to tens of femtoseconds, significantly constraining the direct applicability of pulse-based qubit control schemes. 

To overcome ultrafast relaxation timescales set by the light-matter interaction strength $g$, previous schemes for quantum state control in plasmonic nanocavities with dipole emitters required continuous laser driving \cite{Gonzalez-Tudela2011} or   intense laser pulses with duration $T\gg 1/g$ \cite{Rousseaux2016}. In these schemes, the near field acts both as a channel for driving coherence between internal emitter states and as broadband reservoir inducing decoherence. Alternative to fighting decoherence using strong laser fields, recent work shows that emitter coherence and entanglement lifetimes can be pushed beyond the picosecond barrier by structuring the nanocavity geometry with sub-nanometer resolution to limit the detrimental effects of higher-order plasmonic modes \cite{Crookes2025}. Driving quantum emitters with phase-locked pulse trains could potentially also be used for accumulating population and coherence, but known schemes require the emitter lifetime to be much longer than the pulse repetition time~\cite{Teets1977,Felinto2003,Moreno2014}, which is challenging to achieve in nanocavities.

\begin{figure}[b]
\centering
\includegraphics[width=\columnwidth]{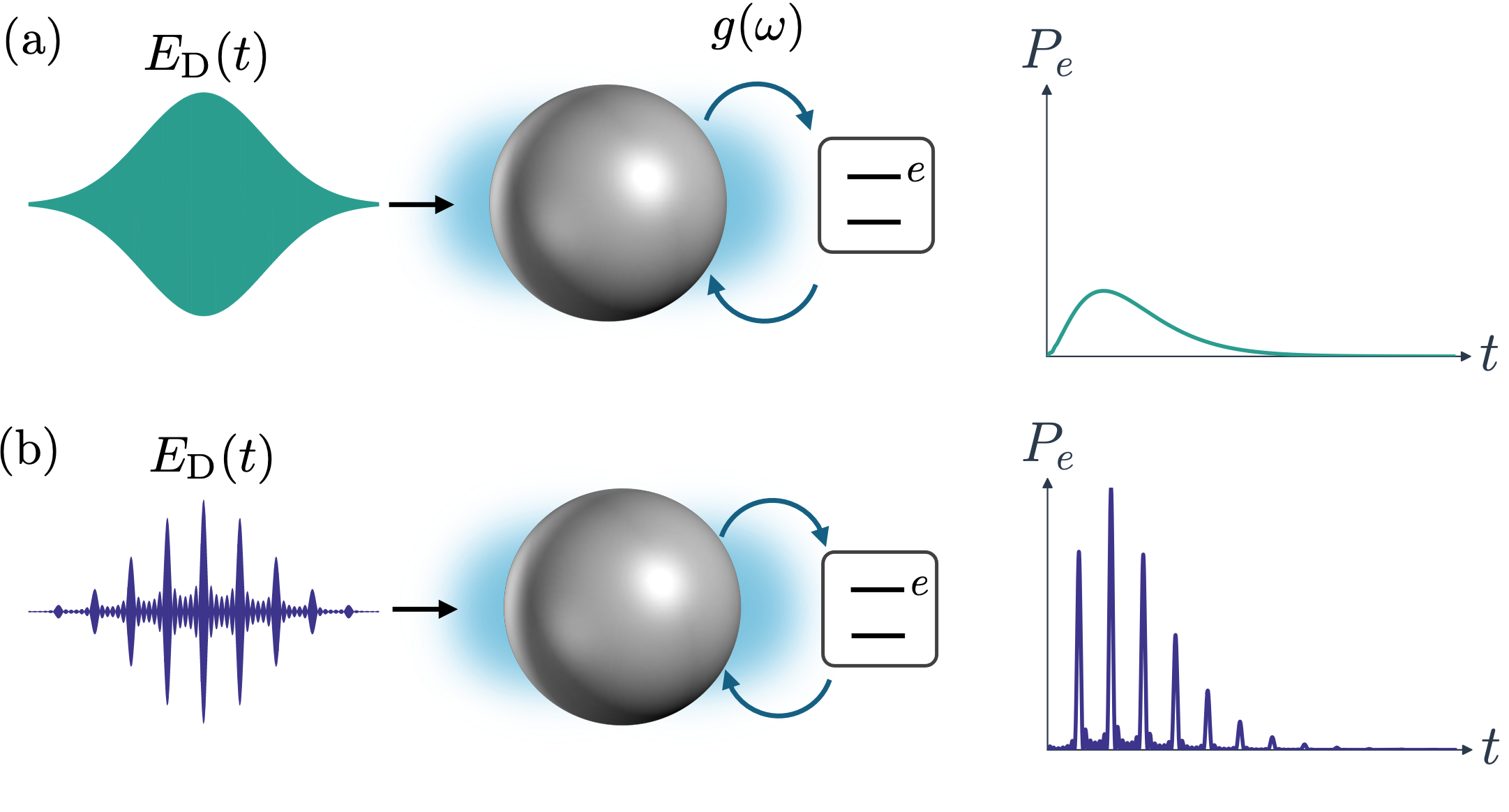}
\caption{{\bf Quantum control scheme}. (a) An emitter embedded in the near field of a spherical nanocavity is driven by a short laser pulse $E_{\rm D}(t)$. The emitter is excited by the near field and undergoes spontaneous emission. (b) Driving the same emitter-nanocavity system with a spectrally-shaped pulse produces a series of recurrent temporally sharp emission events with enhanced amplitude. $g(\omega)$ is the light-matter coupling function.}
\label{fig:scheme}
\end{figure}

In this Letter, we propose a strategy for coherent control of quantum emitters in which the near field of the nanocavity is driven a frequency comb laser. Expanding previous work on driven quantum baths \cite{Reichert2016}, we show that driving the broadband nanocavity near field with a weak frequency comb laser having $N$ phase-locked comb teeth within the nanocavity bandwidth, excited population and coherence in the quantum emitter can be accumulated, provided that the emitter-nanocavity interaction is strong. We derive optimal excitation conditions for inducing a single emission event with a sharp temporal profile of sub-picosecond duration. The generated near-field photon is shown to carry the same spectral profile as the driving frequency comb laser. 

The scheme is illustrated in Fig. \ref{fig:scheme}. Initially the near field is in the vacuum and the emitter in the ground state. Excitation of the near field with a single weak pulse [Fig. \ref{fig:scheme}(a)] creates a nanocavity photon that drives the emitter to an excited state, followed by exponential decay back to the reservoir. {By shaping the laser pulse profile into a train of phase-locked short pulses whose frequency comb spectrum has $N$ components (teeth)} [Fig. \ref{fig:scheme}(b)], quantum interference effects enabled by strong coupling leads to coherent accumulation of emitter population scaling as $N^2$, with a spontaneous emission profile structured into a series of sharp emission events at integer multiples of the recurrence time set by the driving comb spectrum. We show below that these properties can be exploited for engineering nanocavity-based quasi-deterministic single photon sources.  

In order to capture the non-Markovian dynamics of the broadband near-field reservoir strongly coupled to a quantum emitter, we solve the unitary evolution of the coupled light-matter system using macroscopic quantum electrodynamics~\cite{Scheel2008macroscopic,Medina2021,Feist2021,Tudela2024}. We adopt the Lorentzian kernel approximation in Ref.~\cite{Rema2025} for solving the associated integro-differential equations of motion. The two-level emitter with states $\ket{g}$ and $\ket{e}$, transition frequency $\omega_e$, and dipole moment $d_{eg}$ couples to the quantized near field of a plasmonic nanocavity, described by emitter-centered continuum modes $\hat{a}(\omega)$~\cite{Feist2021}. In the rotating-wave approximation, the system Hamiltonian is
$\hat{\mathcal{H}}_{\rm sys} = \hbar\omega_e\ket{e}\!\bra{e}+\int_{0}^{\infty}\!d\omega\,\{\hbar\omega\,\hat{a}^\dagger(\omega)\hat{a}(\omega)+[g(\omega)\,\hat{d}^{(+)}\hat{a}(\omega)+{\rm H.c.}]\}$,
where $\hat{d}^{(+)}=d_{eg}\ket{e}\!\bra{g}$. The coupling function $|g(\omega)|^{2}\propto\mathbf{e}_\alpha\cdot\operatorname{Im}[\mathbf{G}(\mathbf{r}_0,\mathbf{r}_0,\omega)]\cdot\mathbf{e}_\alpha$ is determined by the dyadic Green tensor $\mathbf{G}$ at the emitter position $\mathbf{r}_0$, projected on the dipole orientation $\mathbf{e}_\alpha$~\cite{Scheel2008macroscopic,Feist2021}. The laser drives only the reservoir modes, $\hat{\mathcal{H}}_{\rm D}(t)=\int_{0}^{\infty}d\omega\,[\hbar D(\omega)\,\hat{a}(\omega)\,e^{i\omega t}+{\rm H.c.}]$, so the emitter is excited only through the cavity-filtered near field.

We solve the time-dependent Schr\"odinger equation with the total Hamiltonian $\hat{\mathcal{H}}=\hat{\mathcal{H}}_{\rm sys}+\hat{\mathcal{H}}_{\rm D}(t)$ using the Wigner--Weisskopf ansatz 
$\ket{\psi(t)}=C_{g0}(t)\ket{g}\ket{\{0\}}+C_{e0}(t)e^{-i\omega_e t}\ket{e}\ket{\{0\}}+\int d\omega\,C_{g1}(\omega,t)e^{-i\omega t}\ket{g}\ket{\vb{1}(\omega)}$,
where $C_{g1}(\omega,t)$ is the single-photon wavefunction. We consider weak driving, where the drive enters to linear order in $D(\omega)$ and the single-excitation ansatz preserves the norm~\cite{Rema2025}. Starting from the ground state with the cavity in vacuum, we formally integrate the equation for $C_{g1}(\omega,t)$ and substitute it into the equation for $C_{e0}$. This gives an integro-differential equation for the emitter amplitude,
\begin{equation}\label{eq:IDE}
\dot C_{e0}(t) = -\int_{0}^{t}\!dt'\,\mathcal{K}(t-t')\,C_{e0}(t') + \mathcal{S}(t),
\end{equation}
where $\mathcal{K}(\tau)=\int d\omega\,\mathcal{K}(\omega)\,e^{-i(\omega-\omega_e)\tau}$ is the memory kernel, whose spectrum can be written in terms of the coupling function  as $\mathcal{K}(\omega)=|d_{eg}|^{2}|g(\omega)|^{2}/\hbar^{2}$. The kernel spectrum has bandwidth $\Gamma_K$ (FWHM). The source term is given by
\begin{equation}\label{eq:source}
\mathcal{S}(t) = -\,t\int_{0}^{\infty}\!d\omega\,\sqrt{\mathcal{K}(\omega)}\,D^{*}(\omega)\,e^{-i(\omega-\omega_e)t},
\end{equation}
and encodes the amplitude of the driving field at each frequency $D(\omega)$, filtered by kernel through the function $\sqrt{\mathcal{K}(\omega)}$.

In the weak-driving regime, a single narrowband driving pulse cannot compete with the sub-picosecond plasmonic dissipation. We therefore drive the system with a frequency comb having $N$ phase-locked teeth of equal amplitude and phase, i.e.,  $D(\omega)=D_0\sum_{n=1}^{N}e^{-(\omega-\omega_n)^{2}/2\sigma^{2}}$. Each comb tooth is Gaussian with bandwidth $\sigma$ centered at $\omega_n=\omega_1+(n-1)\Delta\omega$. The teeth span a window of width $\Gamma_K$ centered on $\omega_e$, so that $\Delta\omega=\Gamma_K/(N-1)$. 
We limit the driving strength to $|D_0|\approx800\,\mathrm{Hz}^{1/2}$ to ensure that the system remains within the single-excitation manifold throughout the propagation interval. In the time domain, the frequency comb corresponds to $N$ coincident pulses with the same temporal profile but different relative phases. The phase profile leads to short pulses  separated by $T_{\rm rec}=2\pi\hbar/\Delta\omega$ within an overall Gaussian temporal profile with duration $\sim\hbar/\sigma$ [Fig.~\ref{fig:scheme}(b)]. Physically, $D(\omega)$ describes the laser spectrum filtered by the confined-mode spectrum~\cite{Rema2025}. Because this filtering is linear~\cite{Lamprecht1999,Anderson2010}, a phase-locked far-field pulse train with spectrum within $\Gamma_K$ produces a near-field pulse train with the same period, delayed and broadened only on the plasmon timescale $\sim\hbar/\Gamma_K$.

Fitting $\mathcal{K}(\omega)$ and $\sqrt{\mathcal{K}(\omega)}D^{*}(\omega)$ to sums of Lorentzians makes Eq.~\eqref{eq:IDE} solvable in Laplace space \cite{Rema2025,Rema2026coatings} (see details in End Matter). The emitter response $[s+\tilde{\mathcal{K}}(s)]^{-1}$, where $\tilde{\mathcal{K}}(s)$ is the Laplace transform of the kernel, has drive-independent poles $\{\lambda_k\}$. These are the hybrid emitter--plasmon modes, which form polaritons in strong coupling \cite{Rema2026coatings}. Each fitted comb tooth instead contributes a second-order pole at $-\gamma_n$, with $\gamma_n=\beta_n+i(\eta_n-\omega_e)$ set by its width and center. The emitter excited state amplitude can thus be written as 
\begin{equation}\label{eq:ce0 solution}
C_{e0}(t)=\sum_k c_k e^{\lambda_k t/\hbar}+\sum_n(a_n+b_n t)e^{-\gamma_n t/\hbar}.
\end{equation}
The first summation describes the free response of the hybrid modes, which decays within $T_1\equiv \hbar/|{\rm Re}\,\lambda_k|\sim 50$ fs for typical emitter nanocavity systems in strong coupling \cite{CuarteroGonzalez2020,Rema2025}. The second summation describes the field driven response, whose linear term describes the coherent feeding of the emitter by the comb. The feeding contribution lasts for a timescale $\hbar/\beta_n$. For the type of comb spectrum we consider, we have $\beta_n=k\sigma\sim 1 \,{\rm ps}$ with $k= 0.93$ for all $n$-components (see derivation in Appendix A). 

\begin{figure*}[t]
\centering
\includegraphics[width=\textwidth]{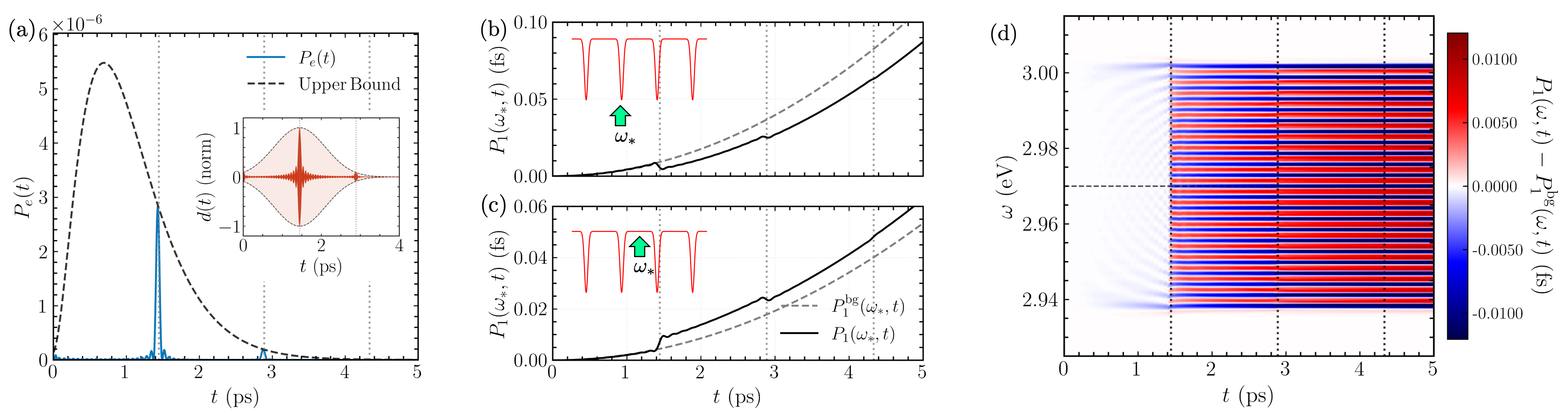}
\caption{{\bf Single-photon emission dynamics.} (a) Excited-state population $P_e(t)$ of dipole emitter (solid line) driven by a frequency comb with temporal profile $E_D(t)$ shown in the inset. Dashed line is the analytical upper bound. (b) Single-photon probability $P_1(\omega^{*},t)$ at the  frequency $\omega_*$ centered at one of the comb teeth (solid line). Dashed line is the driven-cavity background $P_1^{\rm bg }=|D(\omega)|^{2}t^{2}$ (dashed). (c) The same as (b) with $\omega_*$ in between comb teeth. (d) Evolution of the single photon probability density $P_1(\omega,t)$ relative to the driven-cavity background $P_1^{\rm bg}$. The horizontal dashed line marks the emitter frequency $\omega_e=2.97$ eV. The number of frequency comb teeth is $N=23$, the tooth bandwidth is $\sigma=1$~meV, and kernel bandwidth $\Gamma_K=63$ meV. Dotted vertical lines are integer multiples of the recurrence time $T_{\rm rec}$.}
\label{fig:Cg1}
\end{figure*}

Solving for $C_{e0}$ gives the one-photon wavepacket amplitude $C_{g1}(\omega,t)\equiv C_{g1}^{\rm bg}+C_{g1}^{\rm dip}$, where \mbox{$C_{g1}^{\rm bg} = -i D^{*}(\omega)\,t$} is the driven background contribution, and 
\begin{equation}\label{eq:Cg1 dipole}
C_{g1}^{\rm dip} = -i\sqrt{\mathcal{K}(\omega)}\!\int_{0}^{t}\!\!d\tau\,
C_{e0}(\tau)\, e^{i(\omega-\omega_e)\tau}
\end{equation}
{is the emitted contribution from the dipole.} The probability $P_{1}\equiv|C_{g1}(\omega,t)|^2$ of creating a single near-field photon of frequency $\omega$ at time $t$ is therefore given by the background photon injection rate $P^{\rm bg}_{1}\equiv |D(\omega)|^2t^2$, the direct excited dipole emission into the near-field proportional $|C_{g1}^{\rm dip}|^2$, and interference between laser driving and dipole emission $2\,\mathrm{Re}[(C_{g1}^{\rm bg})^{*}C_{g1}^{\rm dip}]$. 

For the coupled system we study a quantum dot emitter ($d_{eg}=24\,\mathrm{D}$) oriented radially at a distance $h=2\,\mathrm{nm}$ from a silver nanosphere of radius $r=20\,\mathrm{nm}$ [Fig.~\ref{fig:scheme}(a)].  Silver is modeled with Drude permittivity $\epsilon_m(\omega)=\epsilon_\infty-\omega_p^2/(\omega^2+i\omega\Gamma)$, where $\omega_p=7.9\,\mathrm{eV}$, $\Gamma=51\,\mathrm{meV}$, and $\epsilon_\infty=6$~\cite{Vanvlack2012}. The kernel spectrum $\mathcal{K}(\omega)$ is obtained using Mie theory for the dyadic Green tensor~\cite{Mie1908,Bohren1983}. The kernel and Lorentzian fits used for solving Eq. (\ref{eq:IDE}) are given in Appendix A.  The emitter frequency $\omega_e=2.97\,\mathrm{eV}$ is resonant with the maximum of $\mathcal{K}(\omega)$, which guarantees strong coupling conditions \cite{Rema2025}. Other nanophotonic setups that support strong coupling would give similar results.

Figure~\ref{fig:Cg1}(a) shows the evolution of excited population of the quantum dot emitter (solid line) subject to a frequency comb drive with $N=23$ teeth within the kernel bandwidth $\Gamma_K=63\,{\rm meV}$. Each teeth has bandwidth parameter $\sigma=1$~meV. These specific driving parameters can be shown to be optimal, i.e., they maximize the excited state population very close to the upper bound  $P_{\rm ub}(t)$ (dashed line, derived in Appendix B), and only one sharp revival event at the time $T_{\rm rec}= 1.44$ ps is significant within the emitter lifetime. The inset shows the pulse train in the time domain. Below we discuss conditions for the optimal number of comb teeth $N_{\rm opt}$. 

The duration of individual emission events is set by the light-matter coupling strength. For the system parameters used here ($h = 2$ nm, $d_{eg}$=24 D; $r= 20$ nm), strong coupling is achieved with Rabi splitting in the far-field spectrum of about $134$ meV \cite{Rema2025}, corresponding in the time-domain to low-$Q$ ultrafast Rabi oscillations having $\sim 30$ fs population exchange timescales with the near field. These ultrafast exchange times achieved in strong coupling allow for individual resolution of emission events. We show in Appendix D that the same emitter driven by the same frequency comb field at larger sphere-emitter distances ($h = 10$ nm) such that strong coupling is lost and weak coupling is established, periodic emission events still occur at the recurrence times, but with emission profiles that have long exponential tails that prevent the resolution of individual events.

Figures~\ref{fig:Cg1}(b) and \ref{fig:Cg1}(c) shows the single-photon probability function $P_1(\omega_*,t)$ associated to the quantum emitter dynamics in Fig. ~\ref{fig:Cg1}(a). The evolution is show at fixed frequency $\omega_*$, chosen to match one of the frequency comb tooth [Fig. \ref{fig:Cg1}(b)] or in between consecutive comb teeth [Fig. \ref{fig:Cg1}(c)]. Other frequency choices show qualitatively the same behavior. Subsequent emission event at integer multiples of $T_{\rm rec}$ modify the photon evolution from the driven-oscillator background growth $P_1^{\rm bg}\sim t^2$ due to dipole interference contributions coming from Eq. (\ref{eq:Cg1 dipole}). At a comb tooth frequency [Fig. \ref{fig:Cg1}(b)], the dipole contribution destructively interferes with the laser-induced amplitude, producing dips in $P_1(\omega_*,t)$ that could in principle be measured using coherent near-field imaging techniques \cite{Park2019}.

Figure~\ref{fig:Cg1}(d) shows the entire near-field single-photon probability density $P_1(\omega,t)$ corresponding to the emission profile in Fig. \ref{fig:Cg1}(a). To highlight the emitter-induced dynamics, the background dipole-free contribution $P_1^{\rm bg}$ is subtracted. Before the first emission burst ($0<t<T_{\rm rec}$), the quasi-instantaneous response of the quantum emitter to the comb-driven near-field mainly leads to small amplitude single-photon interference patters similar to those described in previous work as characteristic ultrafast signatures of strong coupling \cite{Rema2025}. After the first dipole emission event at $T_{\rm rec}$, the frequency comb spectrum becomes imprinted on the near-field single photon probability density. 

As mentioned above, for a comb with $N$ teeth the excitation probability of the quantum emitter can be shown to be upper bounded by 
\begin{equation}\label{eq:bound}
P_{\rm ub}(t)=N^{2}|b|^{2}\,t^{2}e^{-2\beta t/\hbar},
\end{equation}
as derived from Eq. (\ref{eq:ce0 solution}) and $|C_{e0}(t)|\leq \sqrt{P_{\rm up}}$. The bound assumes that linear-in-time state amplitudes $|b_j|t$ dominate the emitter evolution with all terms identical (i.e., $b_n=b$, $\beta_n=\beta$).  (derivation in Appendix B). This expression  describes early time coherent accumulation of excitation scaling quadratically with $N$, consistent with previous work on two-level atoms driven by weak pulse trains \cite{Felinto2001,Felinto2003,Stowe2006}. Equation (\ref{eq:bound}) does not capture saturation effects that can be expected for large $N$ \cite{Tang2008}.

\begin{figure}[t]
\centering
\includegraphics[width=0.45\textwidth]{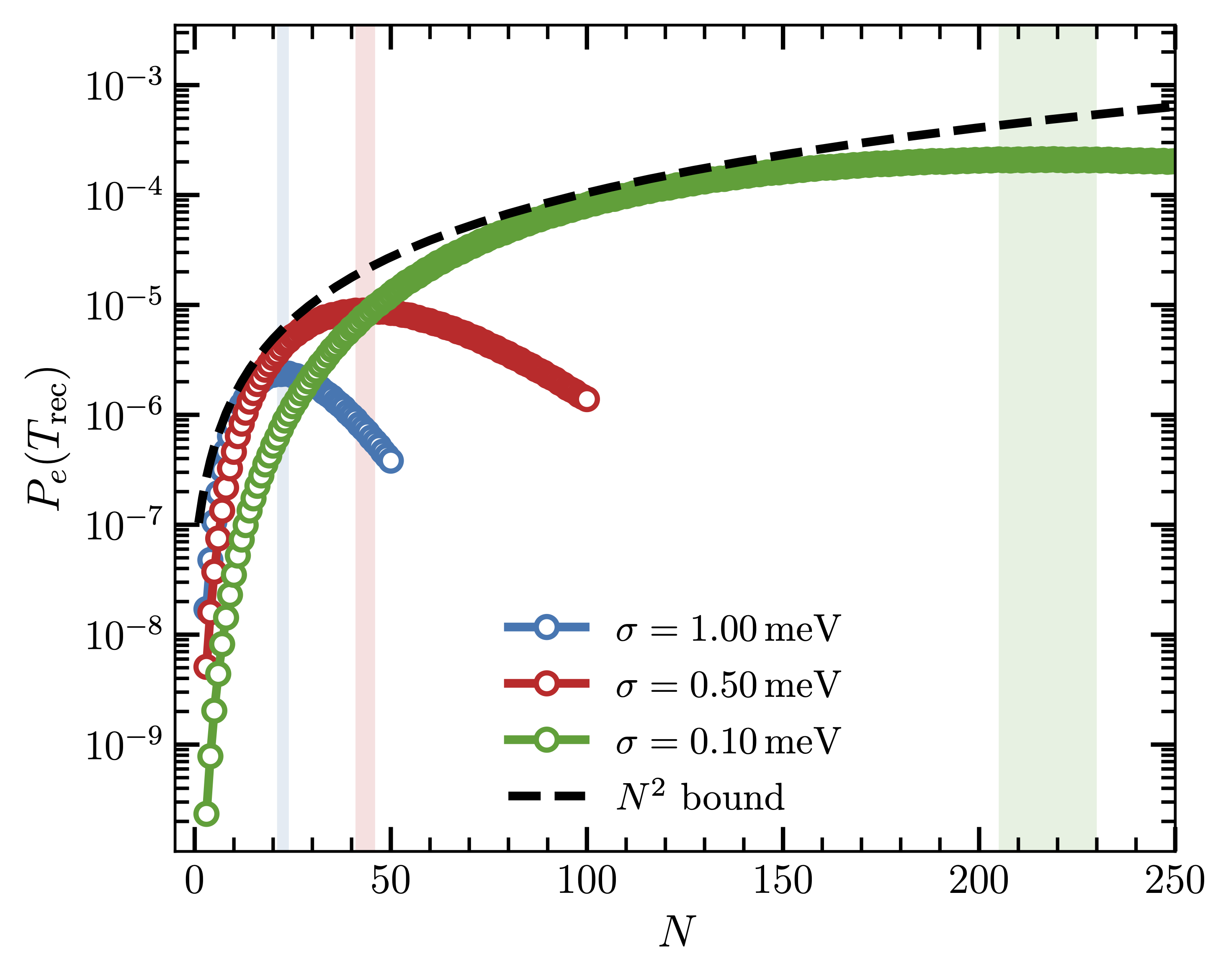}
\caption{{\bf Optimal single-photon pulses.} Emission probability at the first revival time $P_e(T_{\rm rec})$ as a function of the number of comb teeth $N$, for several tooth bandwidth parameters $\sigma$. Dashed line marks the analytical upper bound. Shaded areas mark 1\% variations from the optimal amplitude.}
\label{fig:scaling}
\end{figure}

Figure~\ref{fig:scaling} shows the excited emitter population at the first revival $P_e(T_{\rm rec})$ obtained numerically using Eq. (\ref{eq:ce0 solution}) as a function of $N$, for the same emitter-nanocavity configuration used in Fig. \ref{fig:Cg1}. The frequency comb spacing $\Delta\omega$ is adjusted such that $N$ comb teeth fit within the bandwidth of the nanocavity kernel, i.e., $N =1+\Gamma_K/\Delta\omega$. The peak population at first revival closely follows the coherent accumulation upper bound $P_{\rm ub}$ in Eq. (\ref{eq:bound}). At fixed comb tooth bandwidth $\sigma$, increasing $N$ (reducing $\Delta \omega$) leads to significant deviations from $P_{\rm ub}$ because the revival time $T_{\rm rec}\propto 1/\Delta \omega$ is long enough for the interference between different components in Eq. (\ref{eq:ce0 solution}) to differ from perfect in-phase recurrence assumed to derive Eq. (\ref{eq:bound}). $P_e(T_{\rm rec})$ then reaches a maximum and rolls off with increasing $N$. 

The upper bound  in Eq. (\ref{eq:bound}) suggests that there should be an optimal number of teeth $N_{\rm opt}$ for which the optimal recurrence time $T_{\rm rec}^{\rm opt}\propto N_{\rm opt}$ is just long enough for having the first dipole emission event at $T_{\rm rec}$ with significant amplitude proportional to $N_{\rm opt}^2$, but the probability of subsequent emission events becomes exponentially small, with event rates decaying at the rate $2\beta/\hbar$. It is shown in Appendix C that  $N_{\rm opt}= \Gamma_K/\pi\beta $. The shaded bands in Fig.~\ref{fig:scaling} show peak population values within $1\%$ of the predicted $N_{\rm opt}$. For a quantum emitter driven by an optimal number of comb teeth $N_{\rm opt}$, the probability of the $m$-th population revival event $P_e(m)\equiv P_e(mT_{\rm rec})$ relative to the first event can be written as
\begin{equation}\label{eq:Pofm}
P_e(m)/P_e(1)=\, m^2 {\rm e}^{-2(m-1)\beta T_{\rm rec}/\hbar}. 
\end{equation} 
For example, the second emission event in Fig. \ref{fig:Cg1}(a) at $2T_{\rm rec}\approx 2.9$ ps is only $4e^{-4}\approx 7\%$ as intense as the first event, and the third event has negligible probability ($\sim 0.3\%$).

{\it Conclusions}. We showed that driving the near-field modes of a lossy nanocavity with a weak, phase-locked frequency comb with $N$ teeth enhances the excitation probability of a strongly coupled two-level emitter by a factor scaling  $N^{2}$, due to coherent accumulation processes mediated by the near field that can be understood analytically. For strong coupling with the near field, the excited emitter dipole undergoes rapid emission events over a timescale determined by the light-matter interaction strength. We derive optimal pulse conditions for obtaining an ultrafast quasi-deterministic single photon source that operate at room temperature.

Our methodology is restricted to the single-excitation manifold, thus limiting the validity of the control scheme to weak spectrally-structured laser pulses. Extensions to higher excitation manifolds would allow us to discuss ground-state depletion and multi-photon generation processes, which are relevant for emitter architectures based on organic molecules with rich internal level structures \cite{Wang:2019,Esteso2023,Herrera2022meta}. We illustrate the population control scheme with metallic nanocavities having plasmonic resonances at optical frequencies, but the same physics can be expected for infrared nanocavities \cite{Herrera2020,Wilcken2023,Triana2022}. The pulses needed for implementation of the proposed scheme are available using standard pulse shaping techniques~\cite{Weiner2000}. More broadly, our results show that strong coupling~\cite{GarciaVidal2021} is a necessary condition for coherent control of the coupled emitter-nanocavity system. The nanocavity vacuum field mediates the transfer of coherence from laser to emitter, opening pathways toward nanoscale quantum information processing at room temperature.
\\

\begin{acknowledgments}
A.S.R. is supported by the USACH Excellence Graduate Scholarship. A.E.R.L. is supported by ANID-Fondecyt Iniciaci\'on No.~11250638. F.H. is supported by ANID through Fondecyt Regular Grant No.~1262603. All authors acknowledge support from the ANID-Millennium Science Initiative Program No.~ICN17\_012.
\end{acknowledgments}
\bibliography{SPEcontrol_V1}

\clearpage

\appendix
\section{End Matter}

\subsection{Appendix A: Lorentzian kernel approximation and Laplace solution}
\label{app:solution}

\begin{figure}[h]
\centering
\includegraphics[width=\columnwidth]{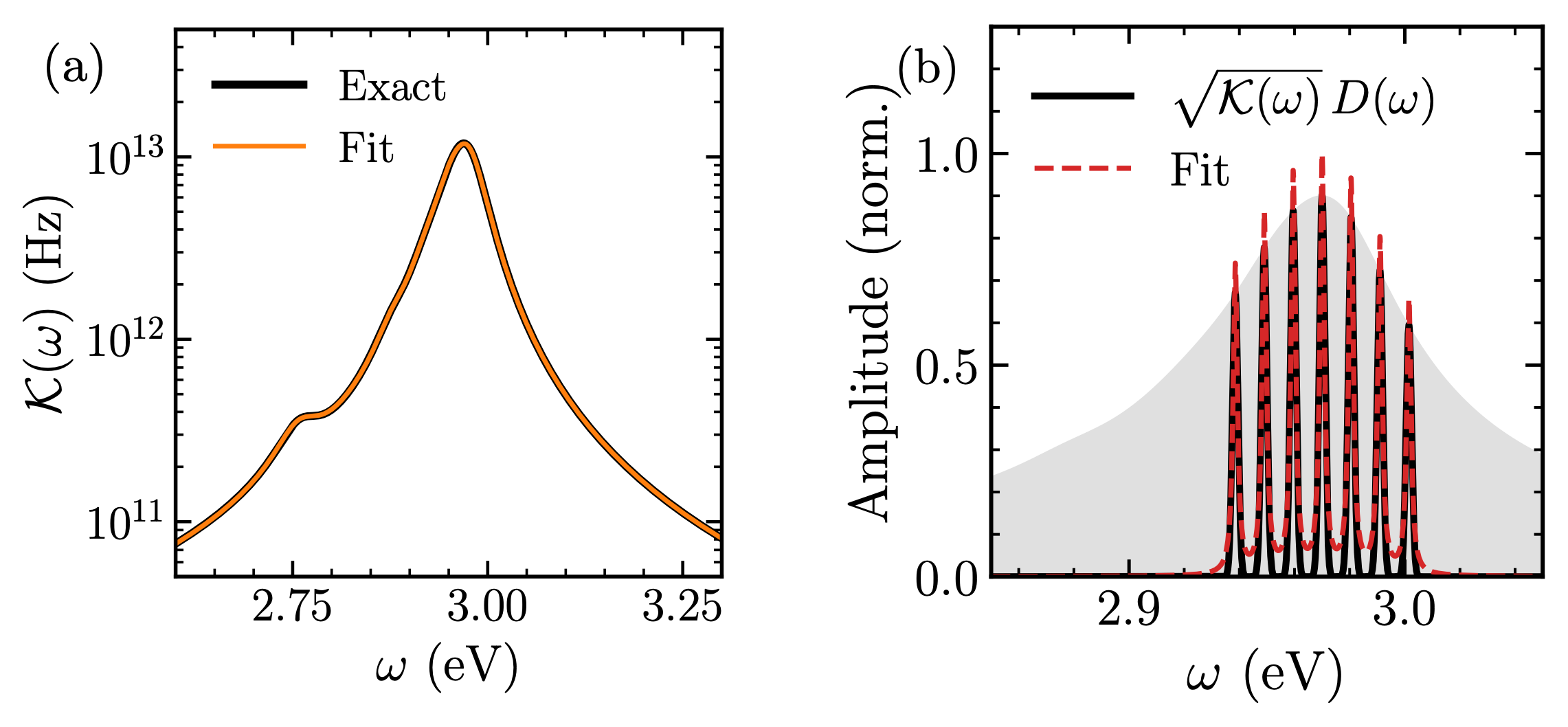}
\caption{Lorentzian fits. (a)~Spectral density $\mathcal{K}(\omega)$ from Mie theory (black) and its six-Lorentzian fit (orange), with parameters in Table~\ref{tab:fit}. (b)~Drive overlap $\sqrt{\mathcal{K}(\omega)}\,D^{*}(\omega)$ for $N=7$ and $\sigma=1$~meV (black) and its fit with one Lorentzian per tooth (dashed red), both normalized to the maximum of the fit. Shaded: $\sqrt{\mathcal{K}(\omega)}$ on the same scale.}
\label{fig:spectrum}
\end{figure}

We express frequencies and rates in energy units, so that time enters through $t/\hbar$. We compute $\mathcal{K}(\omega)$ from the dyadic Green tensor of the sphere using Mie theory and fit it to a sum of $N_L=6$ Lorentzians,
\begin{equation}
\mathcal{K}(\omega)\simeq\frac{1}{\pi}\sum_{j=1}^{N_L}\frac{A_jB_j}{(\omega-\Omega_j)^{2}+B_j^{2}}.
\label{eq:A_Kfit}
\end{equation}
Five components with positive weight describe the plasmon response, and a broad component with negative weight corrects the high-frequency tail [Table~\ref{tab:fit}, Fig.~\ref{fig:spectrum}(a)]. The weights add up to the area of $\mathcal{K}(\omega)$, and $2\sum_jA_j=(97\,\mathrm{meV})^{2}$ exceeds $(\Gamma_K/2)^{2}=(31.5\,\mathrm{meV})^{2}$, so the strong-coupling criterion holds. With $\tilde B_j=B_j+i(\Omega_j-\omega_e)$, the kernel is $\mathcal{K}(t)=\sum_jA_je^{-\tilde B_jt/\hbar}$, with Laplace transform $\tilde{\mathcal{K}}(s)=\sum_j A_j/(s+\tilde B_j)$. The emitter response $G(s)=[s+\tilde{\mathcal{K}}(s)]^{-1}=P(s)/Q(s)$ is then a rational function, with $P(s)=\prod_j(s+\tilde B_j)$ and $Q(s)$ of degree $N_L+1=7$. The roots $\lambda_k$ of $Q(s)$ are the hybrid emitter--plasmon modes. 
\begin{table}[h]
\caption{Lorentzian parameters of $\mathcal{K}(\omega)$ for $h=2$~nm.}
\label{tab:fit}
\begin{ruledtabular}
\begin{tabular}{cccc}
$j$ & $\Omega_j$ (eV) & $B_j$ (eV) & $A_j$ (meV$^2$) \\
\hline
1 & 2.759 & 0.0330 & $60.42$ \\
2 & 2.878 & 0.0232 & $79.27$ \\
3 & 2.926 & 0.0300 & $417.6$ \\
4 & 2.955 & 0.0241 & $1145$ \\
5 & 2.974 & 0.0257 & $3019$ \\
6 & 3.139 & 0.2901 & $-3.39$ \\
\end{tabular}
\end{ruledtabular}
\end{table}

\begin{table}[h]
\caption{Lorentzian parameters of the drive overlap $\sqrt{\mathcal{K}(\omega)}D^{*}(\omega)$ for $h=2$~nm, with $N=7$ teeth of width $\sigma=1$~meV spanning $\Gamma_K$ about $\omega_e=2.97$~eV and $|D_0|=10^{-5}$~a.u. [Fig.~\ref{fig:spectrum}(b)]. The kernel parameters are given in Table~\ref{tab:fit}.}
\label{tab:drivefit}
\begin{ruledtabular}
\begin{tabular}{cccc}
$n$ & $\eta_n$ (eV) & $\beta_n$ (meV) & $u_n$ ($10^{-2}$~meV$^{2}$) \\
\hline
1 & 2.9385 & 0.9254 & $2.719$ \\
2 & 2.9490 & 0.9252 & $3.149$ \\
3 & 2.9595 & 0.9251 & $3.503$ \\
4 & 2.9700 & 0.9249 & $3.649$ \\
5 & 2.9805 & 0.9249 & $3.438$ \\
6 & 2.9910 & 0.9252 & $2.932$ \\
7 & 3.0015 & 0.9254 & $2.409$ \\
\end{tabular}
\end{ruledtabular}
\end{table}

Each tooth of the drive overlap $\sqrt{\mathcal{K}(\omega)}D^{*}(\omega)$ is fitted by least squares with a single Lorentzian of the form of Eq.~\eqref{eq:A_Kfit}, with weight $u_n$, half-width $\beta_n$, and center $\eta_n\simeq\omega_n$ [Fig.~\ref{fig:spectrum}(b), Table~\ref{tab:drivefit}]. {For Gaussian teeth of width $\sigma$ the fitted half-width is $\beta_n = A\sigma$ for every tooth. Measured in units of $\sigma$, the Gaussian is parameter free and the shape of the Lorentzian depends only on $A$. The constant $A$ is thus the Lorentzian-equivalent width of a Gaussian line, set by the two lineshapes rather than by the drive or the nanocavity, and carrying out the fit analytically gives it as the root of
\begin{equation}
e^{A^{2}/2}\,\mathrm{erfc}\!\left(A/\sqrt{2}\right)\left(1+2A^{2}\right)
= 2A\sqrt{2/\pi},
\label{eq:Aroot}
\end{equation}
namely $A \simeq 0.93$. The fits reproduce this value for every tooth and every $\sigma$ used here (Table~\ref{tab:drivefit}). The small differences between teeth come from $\sqrt{\mathcal{K}(\omega)}$ not being flat across each one. All results are computed with this Lorentzian representation.}

The source then becomes $\mathcal{S}(t)=-t\sum_nu_ne^{-\gamma_nt/\hbar}$, with $\gamma_n=\beta_n+i(\eta_n-\omega_e)$, and its transform $\tilde{\mathcal{S}}(s)=-\sum_{n}u_n/(s+\gamma_n)^{2}$ has a second-order pole for each tooth. Inverting $\tilde C_{e0}(s)=G(s)\tilde{\mathcal{S}}(s)$ by residues gives
\begin{equation}
C_{e0}(t)=\sum_k c_k e^{\lambda_k t/\hbar}+\sum_n(a_n+b_n t)\,e^{-\gamma_n t/\hbar},
\label{eq:A_Ce0split}
\end{equation}
with $a_n=-u_nG'(-\gamma_n)$, $b_n=-u_nG(-\gamma_n)/\hbar$, and $c_k=[P(\lambda_k)/Q'(\lambda_k)]\,\tilde{\mathcal{S}}(\lambda_k)$. This is Eq. (\ref{eq:ce0 solution}) in the main text.

Equation~\eqref{eq:IDE} is linear in $D(\omega)$, so all populations scale as $|D_0|^{2}$, and $D_0$ only sets their overall scale. For a comb of $N$ non-overlapping teeth, the injected photon number is $t^{2}\!\int d\omega\,|D(\omega)|^{2}\simeq\sqrt{\pi}\,N|D_0|^{2}\sigma\,t^{2}$, so the single-excitation treatment requires $|D_0|\ll1/(t\sqrt{N\sigma})$, which reduces to the bound of Ref.~\cite{Rema2025} for $N=1$. Our value, $|D_0|=10^{-5}$~a.u. ($\approx800\,\mathrm{Hz}^{1/2}$ for frequencies in units of $\omega/2\pi$), keeps the injected photon number below $5\%$ up to the first burst of every optimal comb: $5\times10^{-4}$, $2\times10^{-3}$, and $0.05$ for $\sigma=1$, $0.5$, and $0.1$~meV.

\subsection{Appendix B: Excitation upper bound and rephasing}
\label{app:dynamics}

Applying the triangle inequality to Eq.~\eqref{eq:A_Ce0split} gives $|C_{e0}(t)|\le E(t)$, with
\begin{equation}
E(t)=\sum_k|c_k|\,e^{{\rm Re}\lambda_k t/\hbar}+\sum_n\big(|a_n|+|b_n|\,t\big)e^{-\beta_n t/\hbar}.
\label{eq:A_full_envelope}
\end{equation}
$E^{2}(t)$ is the dashed curve in Fig.~\ref{fig:Cg1}(a). At long times, $E(t)$ decays at the slowest of the rates $|{\rm Re}\,\lambda_k|$ and $\beta_n$. For Fig.~\ref{fig:Cg1}, $\beta\simeq0.93$~meV lies more than an order of magnitude below the slowest hybrid-mode rate, $12.6$~meV, so the decay is set by the drive.

For $t\gtrsim1$~ps the hybrid-mode terms have decayed and $|a_n|\ll|b_n|t$. This range includes the optimal first bursts for all $\sigma$ considered, $T_{\rm rec}^{\rm opt}=2\hbar/\beta\simeq1.4$--$14$~ps. With a common width $\beta$, the amplitude then factorizes as
\begin{equation}
C_{e0}(t)\simeq t\,e^{-\beta t/\hbar}\,\Sigma(t),\qquad
\Sigma(t)\equiv\sum_{n=1}^{N}b_n\,e^{-i(\eta_n-\omega_e)t/\hbar}.
\label{eq:A_factorized}
\end{equation}
For uniform spacing, the phase factors realign at every multiple of $T_{\rm rec}=2\pi\hbar/\Delta\omega$. Since $|\Sigma(t)|\le\sum_n|b_n|\le N|b|$, with $|b|=\max_n|b_n|$, the population is upper-bounded by
\begin{equation}
P_{\rm ub}(t)=N^{2}|b|^{2}\,t^{2}e^{-2\beta t/\hbar},
\label{eq:A_envelope}
\end{equation}
which is the dashed curve in Fig.~\ref{fig:scaling}. The comb stays below it because $|b_n|=|u_nG(-\gamma_n)|/\hbar$ varies across the band, through both $\sqrt{\mathcal{K}(\omega)}$ and the emitter response. Once the teeth sample the band, this reduction is a constant factor. 

\subsection{Appendix C: Optimal comb and single-photon burst}
\label{app:optimum}

Since $\Sigma(t)$ takes the same value at every $t=mT_{\rm rec}$, the population at the $m$th burst is $P_e(m)=|\Sigma(T_{\rm rec})|^{2}(mT_{\rm rec})^{2}e^{-2m\beta T_{\rm rec}/\hbar}$, and successive bursts differ only through the factor $t^{2}e^{-2\beta t/\hbar}$,
\begin{equation}
\frac{P_e(m)}{P_e(1)}=m^{2}e^{-2(m-1)x},\qquad x\equiv\beta T_{\rm rec}/\hbar.
\label{eq:A_contrast}
\end{equation}
For $x>\ln2$ the first burst is the largest, so the peak population is $P_e(1)$. Increasing $N$ increases $|\Sigma(T_{\rm rec})|\propto N$, but at
fixed span it also delays the first burst, since $N-1=\Gamma_KT_{\rm rec}/(2\pi\hbar)$. For $N\gg1$ this gives
\begin{equation}
P_e(1)\propto x^{4}e^{-2x},
\label{eq:A_Pmax_general}
\end{equation}
which grows as $N^{4}$ for $x\ll1$ and is maximal at $x=2$. Since $x=2>\ln2$, this is also the maximum of the peak population. 
At this optimum, $P_e(2)/P_e(1)=4e^{-4}=7.3\%$ and $P_e(3)/P_e(1)=9e^{-8}=0.3\%$, so the first burst carries $93\%$ of the summed burst peaks. For the comb of Fig.~\ref{fig:Cg1} ($N=23$, $x=2.03$), Eq.~\eqref{eq:A_contrast} gives $P_e(2)/P_e(1)=1/14.5$, in agreement with the
numerics.

\subsection{Appendix D: Weak-coupling limit}
\label{app:weak}
 
\begin{figure}[h]
\centering
\includegraphics[width=\columnwidth]{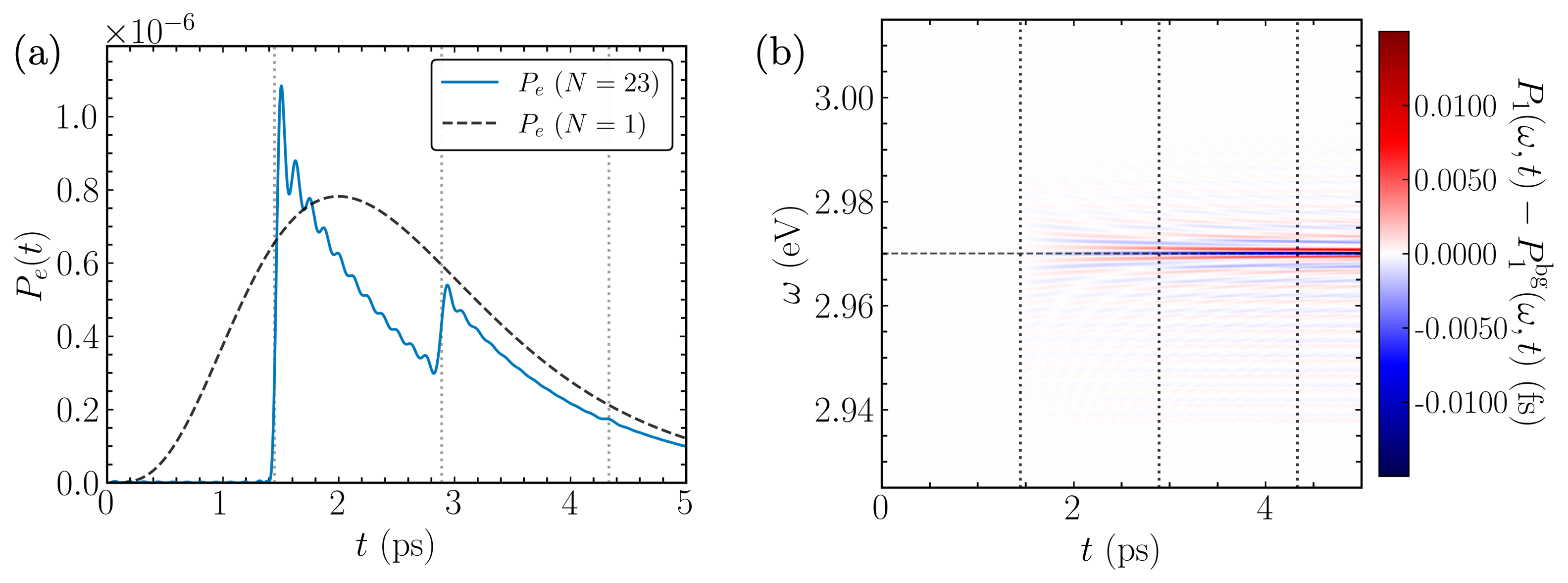}
\caption{{\bf Weak-coupling limit at $h=10$~nm.} Same emitter and comb as in Fig.~\ref{fig:Cg1}. (a) Excited-state population $P_e(t)$ for $N=23$ and for a single tooth of the same amplitude. (b) Emitter-induced change in the single-photon spectrum, $P_1(\omega,t)-P_1^{\rm bg}(\omega,t)$, for the $N=23$ drive. Dotted lines mark $t=mT_{\rm rec}$.}
\label{fig:weak}
\end{figure}
 
At $h=10$~nm from the same sphere, $\mathcal{K}(\omega)$ weakens by two orders of magnitude, so the strong-coupling criterion fails~\cite{Rema2025}. Everything else is left as in Fig.~\ref{fig:Cg1} ($\omega_e=2.97$~eV, $N=23$, $\sigma=1$~meV, span $63$~meV, same $|D_0|$), and $N=23$ remains the optimal comb.
 
The recurrences survive [Fig.~\ref{fig:weak}] with excitations at $t=mT_{\rm rec}$, because their timing is set by the drive and not by the coupling. The amplification does not. A single tooth now excites the emitter almost as strongly as the comb, $P_e^{\max}(N{=}23)/P_e^{\max}(N{=}1)=1.4$ against $317$ at $h=2$~nm, and each burst is followed by a $1.24$~ps exponential tail instead of decaying within its own width. Both follow from the emitter coherence $\hbar/|{\rm Re}\,\lambda_k|$, which is $2.5$~ps here instead of $52$~fs. At $h=2$~nm it is $28$ times shorter than $T_{\rm rec}=1.44$~ps. Here it is longer, so a single tooth has time to excite the emitter on its own. 

The photon spectrum tells the same story [Fig.~\ref{fig:weak}(b)]. The emitter still changes the field, but its response collapses towards $\omega_e$. Nearly half of $|P_1-P_1^{\rm bg}|$ lies within $2$~meV of the emitter frequency, against the $63$~meV spread of Fig.~\ref{fig:Cg1}(d), and it falls by two orders of magnitude within three teeth of $\omega_e$, so the tooth-by-tooth alternation of Figs.~\ref{fig:Cg1}(b) and \ref{fig:Cg1}(c) is lost. The extinction at $\omega_e$ also deepens steadily across the whole window, with only weak steps at $t=mT_{\rm rec}$, so the emission is no longer gated into bursts. The recurrences are therefore a property of the drive, while the amplification is a signature of strong coupling.

\end{document}